\documentclass{emulateapj}
\usepackage{epsf}
\usepackage{apjfonts}
\usepackage{xspace}
\usepackage{amsmath}
   \usepackage{graphicx}

\newcommand{\angstrom}{\textup{\AA}}

\def\fcoll{f_{\rm coll}}

\def\fesc{f_{\rm esc}}
\def\fesca{f_{\rm esc, rel, 0}}
\def\rfesc{f_{\rm esc, rel}}

\def\mmin{M_{\rm min}}
\def\Mcrit{M_{\rm crit}}

\def\Msun{{\rm M}_\odot}

\def\ss{\sigma^2}

\def\NgH{N_{\rm \gamma / H}}
\def\dotNgH{\dot{N}_{\rm \gamma / H}}

\def\dim#1{\mbox{\,#1}}

\def\hide#1{}

\begin{document}

\title{Effect of halo bias and Lyman Limit Systems on the history of cosmic reionization}

\author{Alexander A.\ Kaurov\altaffilmark{1} and Nickolay Y.\ Gnedin\altaffilmark{1,2,3}}
\altaffiltext{1}{Department of Astronomy \& Astrophysics, The
  University of Chicago, Chicago, IL 60637 USA; kaurov@uchicago.edu}
\altaffiltext{2}{Particle Astrophysics Center, 
Fermi National Accelerator Laboratory, Batavia, IL 60510, USA; gnedin@fnal.gov}
\altaffiltext{3}{Kavli Institute for Cosmological Physics and Enrico
  Fermi Institute, The University of Chicago, Chicago, IL 60637 USA} 

\begin{abstract}
We extend the existing analytical model of reionization by \citet{Furlanetto2004} to include the biasing of reionization sources and additional absorption by Lyman Limit systems. Our model is, by construction, consistent with the observed evolution of the galaxy luminosity function at $z\la8$ and with the observed evolution of Ly-$\alpha$ forest at $z\la6$. We also find that, for a wide range of values for the relative escape fraction that we consider reasonable, and which are consistent with the observational constraints on the relative escape fraction from lower redshifts, our reionization model is consistent with the WMAP constraint on the Thompson optical depth and with the SPT and EDGES constraints on the duration of reionization. We, therefore, conclude that it is possible to develop physically realistic models of reionization that are consistent with all existing observational constraints.
\end{abstract}


\section{Introduction}
\label{sec:intro}

Cosmic "Dark Ages" began when the Universe was only 380,000 years old, at the epoch of recombination, when cosmic plasma and radiation have cooled enough to allow electrons to combine with protons to form stable neutral hydrogen atoms. Lasting for some 200 million years, the Dark Ages encompass time when the dark matter was collapsing into bound objects (halos) due to gravity, but no stars were formed yet. The epoch of reionization (EoR) started with the formation of the first star-forming galaxies and quasars. According to current understanding of reionization, the ultra-violet radiation from these objects was the primary source of energy that stripped electrons off the hydrogen atoms, leaving most of intergalactic matter highly ionized. 

There are three primary observational probes that provide information about the EoR. The first one is the Lyman-alpha forest, which can trace the distribution of neutral hydrogen in the universe and the small-scale structure of ionized bubbles, but only up to $z\sim6$ (i.e.\ only late stages of the EoR can be probed with this method). Another probe is the surveys of Lyman-alpha emitters, which can statistically constraint the size distribution of ionized bubbles from the clustering strength of Lyman-alpha-bright galaxies. Finally, Thomson scattering of CMB photons off the free electrons produced during the reionization, measured by WMAP and SPT, can be used to place global, integral constraints on the overall timing and duration of the reionization epoch.

The most promising type of observation, that should become possible in the near future, is detection of the 21 cm transition of neutral hydrogen. The 21 cm emission is a direct probe of the large-scale distribution of neutral hydrogen during the reionization era, and is, therefore, a complementary probe to the Lyman-alpha forest at lower redshifts. Hence, theoretical studies of EoR are timely and relevant.

The epoch of reionization can be studied theoretically with numerical, semi-numerical, and analytical models. Until recently, numerical simulations of the EoR have been limited to small volumes or low resolution \citep{Trac2011}. Besides that, simulations are way too computationally expensive to be used in Markov Chain Monte Carlo (MCMC) analysis for obtaining constraints on cosmological parameters or for producing multiple mock data sets for the future observations. Here semi-numerical and analytical models come into play.

As the mass in the collapsed objects in the universe was growing, the total number of ionizing photons was increasing too. At some point before $z\sim6$, the number of photons became sufficient to ionize all of intergalactic hydrogen. Direct calculation of the number of ionizing photons and intergalactic baryons allows one to track the global ionized fraction \citep[c.f.][and referenced therein]{Kuhlen2012}. However, all the information about morphology of ionized regions is completely lost in such straightforward calculations. A more detailed treatment of this process can be achieved with the analytical model based on the Press-Schechter like formalism, pioneered by \citet[][hereafter FZH04]{Furlanetto2004}. This model allows one to track the bubble size distribution and the mean free path of ionizing photons, as well as the power spectrum of ionized regions, which can be used for calculating secondary anisotropies of the CMB and for the expected 21 cm signal. In this paper we present a logical extension of this analytical model that accounts for the biasing of the ionizing sources and additional absorptions by Lyman Limit systems.

The paper is organized as follows. In \S \ref{sec:model} the brief outline of the FZH04 model is presented. In \S \ref{sec:ourmethod} we describe the key steps of our method, with the details of our implementation to follow. In \S \ref{sec:results} the results are presented and we discuss the benefits and possible applications of our model. And in \S \ref{sec:conclusions} we conclude.

\section{The FZH04 model}
\label{sec:model}

The FZH04 model is based on the idea of calculating the total number of emitted photons in some large enough region of space with known over-density and comparing this number with the number of baryons in that region. If $m_{\rm coll}$ is the mass of the collapsed objects and $m$ is the mass of some region of space, the condition for the region to be ionized is formulated as:

\begin{equation}
\label{eq:zeta}
m=\zeta m_{\rm coll} .
\end{equation}

We can interpret $\zeta$ as ratio of the total number of hydrogen ionizing photons produced by one collapsed hydrogen atom $N_{\gamma/{\rm c}}$ to the number of photons required to ionize one hydrogen atom $N_{i/{\rm H}}$ (which would be 1 if there are no recombinations),
\[
  \zeta = \frac{N_{\gamma/{\rm c}}}{N_{i/{\rm H}}}.
\]

In the presence of recombinations,
\[
  N_{i/{\rm H}}(t) = 1 + \int_0^t \frac{dt}{\bar{t}_{\rm rec}},
\]
where $\bar{t}_{\rm rec}$ is the average recombination time that depends, among other factors, on the clumping factor of the gas. Both, $N_{\gamma/{\rm c}}$ and $N_{i/{\rm H}}$ can, in principle, vary in space. Notice that the gas clumping is only important in the regime when recombinations dominate over the first ionization. In this paper we follow FZH04 and neglect recombinations.

Now let's consider some region in space, and neglect recombinations for now. It is ionized if the number of hydrogen ionizing photons in it is greater than the number of hydrogen atoms times $N_{\gamma/{\rm c}}$.  In order to compare these numbers, we first have to find $m_{\rm coll}$, which is the function of local over-density.  It can be done using, for example, Press-Schechter model \citep{PressSchetcher1974, Bond1991, Lacey1993}, 

\begin{equation}
\label{eq:fcollPS}f_{\rm coll}=\mathrm{erfc}\left[\dfrac{\delta_{\rm c}(z)-\delta_m}{\sqrt{2[\sigma^2(\mmin)-\sigma^2(m)}}\right] .
\end{equation}
where $\sigma^2(m)$ is the variance of density fluctuations on the scale $m$ and $\mmin$ is the minimum mass of an ionizing source (we discuss this mass later in \S \ref{sub:escape})\footnote{Hereafter we adopt the FZH04 notation of using a lower-case $m$ to denote the mass of an ionized region; we also use a standard convention of adopting an upper-case $M$ for the masses of dark matter halos and galaxies. That makes our notation somewhat non-intuitive, but we consider that preferable to using a more logical but totally unfamiliar notation of labeling the masses of ionized regions with $M$ and masses of galaxies with $m$.}. Hereafter we chose convention (adopted by FZH04) in which $\sigma(m)$ is evaluated at the reference redshift $z_*=0$ and $\delta_c(z)$ is a function of redshift. Such convention is convenient because $\delta_c(z)$ becomes the only quantity that explicitly depends on time. 

Hence, the condition that the bubble of mass $m$ at redshift $z$  with over-density $\delta_m$ is ionized becomes

\begin{equation}
\label{eq:collfrac}
\dfrac{1}{\zeta} \leq f_{\rm coll}.
\end{equation}

Equation (\ref{eq:collfrac}) can be rewritten as a constraint on local density:
\begin{equation}
\label{eq:barrier}
\delta_m \geq \delta_x(m,z) \equiv \delta_{\rm c}(z) - \sqrt{2}K(\zeta)\left[ 
\sigma^2(\mmin)-\sigma^2(m) \right]^{1/2},
\end{equation}
where $K(\zeta)=\mathrm{erf}^{-1}(1-\zeta^{-1})$. Using the excursion set formalism and the function $\delta_x(m,z)$ as a barrier, one can find the distribution of the masses (and scales) of ionized
regions. 

More detailed description of this method can be found in FZH04. In the rest of this paper, we assume that the reader is familiar with the excursion set formalism, the concept of barrier crossing, and Monte-Carlo sampling of possible random-walk trajectories as a method to compute the first barrier crossing.

\section{Our approach}
\label{sec:ourmethod}

The method described in the original work by \cite{Furlanetto2004} assumes the constant photon per collapsed baryon ratio. Later in \cite{Furlanetto2006} it was extended by making $\zeta$ a function  of halo mass.  However, the model based on excursion set formalism allows one to  take into account the efficiency of photon production as a function of three parameters: redshift, local over-density, and the bubble size. Below we give a general overview of our approach step by step; it is followed by detailed explanation of each component.

{\bf 1.} In the first step we find the UV luminosity-to-mass ratio of galaxies (i.e.\ ionizing sources) as a function of redshift. One can do that by using abundance matching between the halo mass function \citep{Tinker2008} and the observed luminosity functions of high redshift galaxies \citep{Bouwens2007,Bouwens2010}.  The main uncertainty in this step is the extrapolation of the abundance-matched luminosity-to-mass ratio to higher redshifts, since observations only cover the redshift range $z\lesssim8$, while reionization starts at earlier times. In \S \ref{sub:MLrelation} we describe in details the abundance matching process and our method of extrapolation to higher redshifts that is based only on the theoretically computable evolution of the halo mass function, without any additional parameters.

{\bf 2.} The observations provide us with the information about the galaxy UV luminosity at the wavelength $\sim1600\angstrom$. But the ionizing radiation has the wavelengths below $912\angstrom$. To convert one into another we need to know the typical spectrum of the galaxy and the relative escape fraction of the photons at these two wavelengths. For the typical spectrum we used Starburst99 \citep{Leitherer1999, Vazquez2005, Leitherer2010} and found that ratio of fluxes at $\lambda<912\angstrom$ and $\lambda=1600\angstrom$ is $r_{int}=0.241$ (see Appendix for details). For the escape fraction, we use a simple, two parameters model which is discussed in \S \ref{sub:escape}. Combining these two factors we can rewrite luminosity as:
\[
L_{<912\angstrom}(M,z)=r_{\rm int}\, \rfesc(M)\, L_{1600\angstrom}(M,z)
\]

{\bf 3.} In this step we calculate the number of ionizing photons. In FZH04 the authors used the parameter $\zeta$, which measures the ratio of ionizing photons to collapsed hydrogen atoms. We find it more convenient to use another parameter, $N_{\gamma/H}$, the ratio of the number of hydrogen ionizing photons to the total number of hydrogen atoms in the region (these two parametrization are related to each other by the collapsed fraction). The luminosity-to-mass ratio for ionizing sources, calculated in the first step, allows us to find $\dotNgH$ at the mean cosmic density as a function of redshift, 
\begin{equation}
\dot{N}_{\gamma/H}(z)=\dfrac{\int L_{<912\AA}(M,z)\dfrac{dn}{dM}(z)\, dM}{n_{\rm H,0}},\label{eq:NgammaH}
\end{equation}
where $dn/dM$ is the halo mass function and where $n_{\rm H,0}$ is the mean number density of hydrogen atoms,
\begin{equation}
n_{\rm H}=\dfrac{\Omega_{\rm b}}{\Omega_{\rm m}}(1-Y_{p})\frac{\rho_{\rm crit}}{m_{\rm p}}.
\label{eq:nH}
\end{equation}

At this point we need to account for the halo bias. In over-dense as well as in under-dense regions the halo mass function is different, which makes $\dot{N}_{\gamma/H}$ a function of local over-density $\delta$. Details of how the bias is implemented in our calculations are provided in \S\ref{sub:bias}. Another important factor is the absorption of ionizing radiation by Lyman Limit systems, which are not modeled in the FZH04 formalism and need to be included as a separate component. This absorption depends on the mean free path of an ionizing photon at a given redshift and the size distribution of ionized bubbles. In \S\ref{sub:lls} we describe in details what model for Lyman Limit systems we use and how we calculate  the fraction $f_{LLS}(z,R)$ of ionizing photons that can reach the boundary  of an ionized bubble of size $R$ (in comoving units) and hit a neutral hydrogen atom. Taking the bias and Lyman Limit systems effects into account, Equation (\ref{eq:NgammaH}) becomes
\begin{equation}
\begin{split}
\dotNgH(z,\delta,R) & = f_{\rm LLS}(z,R)\times \\
& \dfrac{\int L_{<912\angstrom}(M,z)\dfrac{dn}{dM}(z,\,\delta,\,R)\, dM}{n_{\rm H,0}(1+\delta)},\label{NgammaH2}
\end{split}
\end{equation}
where $\dfrac{dn}{dM}(z,\,\delta,\,R)$ is a halo mass function at overdensity $\delta$ on scale $R$.
%
%

{\bf 4.} In the most general case $\dot{N}_{\gamma/H}$ is the function of redshift, local over-density, and scale (bubble size). The bubble of size $R$ at some redshift $z$ with some over-density $\delta$ is ionized when $\NgH(z,\delta,R)$ reaches the fraction of non-collapsed hydrogen atoms,
\[
\NgH(z,\delta,R)=1-\fcoll,
\]
where $\fcoll$ is taken from Equation (\ref{eq:fcollPS}) with $\mmin$ being the minimum mass of a
halo that is capable of retaining its gas after (local) reionization
\citep{dsh:hygs06,dsh:ogt08}. In this paper we take that mass to be $10^{8}M_{\odot}$, since the minimum mass of a
halo capable of retaining its gas has not yet been calibrated as a function of local conditions; however, in the future the model can be improved by adopting it as function of redshift and other parameters. This correction is not particularly important, though, since $\fcoll$ is expected to be small.

{\bf 5.} The last step in our model is computing $\NgH$. In the original FZH04 model $\zeta$ was constant and $\NgH$ was a function of $z$ only , hence Equation (\ref{NgammaH2}) could be integrated directly. In the case when $\dotNgH$ depends on $R$ (or, equivalently, on $m$),
$\delta$, and $z$, it is no longer valid to directly integrate Equation (\ref{NgammaH2}) because both $\delta$ and $R$/$m$ can be functions of time. In addition, ionized bubbles can merge, thus combining two separate bubbles with $m_1$ and $m_2$ into a single bubble of mass $m=m_1+m_2$.

In order to account for both of these effects, we need to find the barrier at
time $t+\Delta t$ from the information about the bubble distribution at time $t$. We describe our approach in \S\ref{sub:merging}.

\subsection{Luminosity-Mass Relation}
\label{sub:MLrelation}

For a given galaxy, the production rate of ionizing photons can be
derived from the observed UV luminosity if the intrinsic (i.e.\
unabsorbed) spectrum of the galaxy $L^{\rm int}_{\lambda}$ is known,
\begin{eqnarray}
  \dot{N}_{\gamma} & = & \dfrac{L(<912\angstrom)}{\left\langle
  E\right\rangle_{\rm ion}} \nonumber\\
  & = & \fesc(<912\angstrom)
  \dfrac{L^{\rm int}(<912\angstrom)}{\left\langle E\right\rangle_{\rm ion}} \nonumber\\
  & = & \fesc(<912\angstrom) 
  \dfrac{L^{\rm int}(<912\angstrom)}{\lambda L^{\rm int}_\lambda(1600\angstrom)}
  \dfrac{\lambda L^{\rm int}_\lambda(1600\angstrom)}{\left\langle
  E\right\rangle_{\rm ion}} \nonumber\\
  & = & \fesc(<912\angstrom) 
  \dfrac{L^{\rm int}(<912\angstrom)}{\lambda L^{\rm int}_\lambda(1600\angstrom)}
  \dfrac{\lambda L^{\rm obs}_\lambda(1600\angstrom)}{\fesc(1600\angstrom)
  \left\langle E\right\rangle_{\rm ion}} \nonumber\\
  & = & \rfesc \dfrac{r_{\rm int}}{\left\langle E\right\rangle_{\rm ion}}
  \lambda L^{\rm obs}_\lambda(1600\angstrom),
  \label{eq:dotNgam}
\end{eqnarray}
where $\left\langle E\right\rangle_{\rm ion}$ is the average energy of a
hydrogen ionizing photon, $r_{\rm int}=(L^{\rm int}_{<912\angstrom}/\lambda
L^{\rm int}_\lambda(1600\angstrom)$ is the intrinsic ratio of the ionizing
to UV luminosity, and $\rfesc =
\fesc(<912\angstrom)/\fesc(1600\angstrom)$ is the relative escape
fraction \citep[c.f.][]{ng:gkc08}.

\begin{figure}
\includegraphics[width=1\columnwidth]{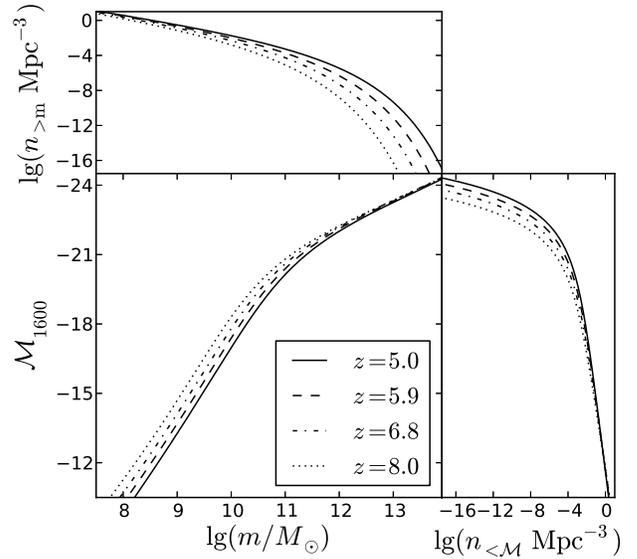}
\label{fig:lmrel}
\caption{Illustration of using the abundance matching to derive the
luminosity-mass relation.  Cumulative mass function of dark matter
halos (top panel) and the cumulative UV luminosity function of
galaxies (right panel) can be matched to obtain the relationship
between the UV absolute magnitude and the halo mass (central panel).
}
\end{figure}

We discuss our model for the relative escape fraction below, in \S
\ref{sub:escape}. For our choice of the intrinsic galaxy spectrum
(described in the Appendix) $r_{\rm int}=0.241$ and $\left\langle
E\right\rangle_{\rm ion}=1.93{\rm Ry}$.

At redshifts where the galaxy luminosity function is actually measured,
all we need to do is to integrate Equation (\ref{eq:dotNgam}) over all
galaxies. However, in Equation (\ref{NgammaH2}) we
need to know $\dot{N}_{\gamma}$ at all redshifts, between the
observational windows and beyond the highest redshift ($z\approx8$) at
which the reliable observational determination of the galaxy
luminosity function exists.

In order to extrapolate beyond $z\approx8$, one can use several approaches. One can, for example, fit the observed evolution of the galaxy luminosity function with some parametric form and extrapolate that parametric form mathematically, like in \citet{Kuhlen2012}. A
limitation of such approach is that any extrapolation depends on
the exact mathematical form of the fitting function; different fitting
functions will produce different extrapolations, thus biasing the
result in an unpredictable way.

It would be preferable to base the extrapolation on a physical
model. In the Press-Schechter framework, the only physical component
is the evolution of the halo mass function. Thus, a physically-based
method of extrapolation is to assign galaxies to dark matter halos and
use the evolving halo mass function to compute the evolving galaxy
luminosity function. 

To find the relation between the halo mass and the galaxy UV
luminosity one can adopt a widely used method of abundance matching:
given the cumulative halo mass function and the cumulative galaxy
luminosity functions, halos of a particular mass $M$ (for some mass
definition, not necessarily the virial mass) can be identified with
galaxies of some luminosity that have the same number density. This
matching does not have to be fully deterministic - scatter can be
included in the relation as long as we treat the abundance-matched
luminosity $L(M)$ as the \emph{average} luminosity of all
galaxies that live in halos of mass $M$.

Such an approach has been used, for example, by \citet{ng:g08} and
\citet{ng:vg09}, and we show an updated version of abundance matching
between the galaxy UV luminosity and the halo virial mass in Figure
\ref{fig:lmrel} for four values of cosmological redshift.

However, as is apparent from Figure
\ref{fig:lmrel}, the luminosity-mass relations obtained by abundance
matching are redshift-dependent. Hence, any extension to higher
redshifts will require extrapolation and will again be subject to
unknown biases.

\citet{rei:tsbo10} circumvented that limitation by matching luminosity
functions not with the actual halo mass functions, but with the
difference between the two halo mass functions at two redshifts
separated by a fixed time interval of $200\dim{Myr}$. With such
matching, the luminosity-mass relations become redshift-independent.

In this paper we propose another method of extrapolation. Our main
motivation is to find a physically plausible relation between the
galaxy luminosity and some property of a dark matter halo, and then
match that property to the halo mass. One such property may be mass
within a fixed proper density threshold. The cosmic overdensity
$\Delta$ changes simply due to the expansion of the universe, while
stars presumably form in the interstellar medium of galaxies at
densities that do not depend on cosmic expansion. 

Hence, if $M_\Delta$ is the mass within the radius enclosing the
cosmic overdensity $\Delta$ (so that $M_{200}$ is the virial mass at
high redshift), we can obtain a relation between the galaxy luminosity
$L$ and $M_\Delta$ by abundance matching,
\[
  n(>L) = n(>M_\Delta),
\]
and choose $\Delta$ to correspond to a given \emph{physical} density
$\Delta_{\rm phys}$,
\[
  \Delta(z) = \frac{\Delta_{\rm phys}}{(1+z)^3}.
\]
For our fiducial value we choose $\Delta_{\rm phys}=6.9\times10^5$
($n_b\approx 0.2\dim{cm}^{-3}$), which corresponds to
the cosmic overdensity $\Delta=3200$ at $z=5$ and overdensity $\Delta=200$ $z=14$.

With that choice of the matched halo mass, the luminosity-mass relation becomes redshift independent - i.e., if we use $M_{\Delta_{\rm phys}}$ in Figure \ref{fig:lmrel} instead of $M_{200}$, all four lines fall on top of each other (we do not show such a figure since it is trivially redundant).

Given the relation between $L$ and $M_\Delta$,
we can derive $L(M_{\rm vir},z)$ at any redshift by abundance-matching
of two theoretical mass functions, $n(>M_\Delta)$ and $n(>M_{\rm
  vir})$, which are known at any redshift, i.e.
\[
  L(M_{\rm vir},z) \equiv L\left(M_\Delta(M_{\rm vir},z)\right).
\]
This procedure does not
involve any extrapolation in time, and, hence, all time evolution
comes through the physical evolution of the halo mass function.

\subsection{Escape Fraction}
\label{sub:escape}

The most sensitive parameter for any reionization model is the escape
fraction of ionizing photons from galaxies. Different definitions of
$\fesc$ exist in the literature. When the actual ionizing emissivity
of all stars in a galaxy is known, $\fesc$ stands for a fraction of
the radiation that leaves the galaxy - it is called the
\emph{absolute} escape fraction then. In reality, the total ionizing flux from
stars in high redshift galaxies is unknown. The observed quantity is the UV luminosity (at
wavelength $\lambda\approx1600\angstrom$) of a galaxy, and, hence, we
are interested not in the absolute, but in the \emph{relative} escape fraction,
$\rfesc = \fesc(<912\angstrom)/\fesc(1600\angstrom$) that
enters Equation (\ref{eq:dotNgam}).

A large body of work exists on modeling the escape fractions from high
redshift galaxies \citep[see, e.g.][for the latest
models]{Razoumov2010,Srbinovsky2010,Wyithe2010,Fernandez2011,Mitra2012}.
In this paper we adopt the second-to-the-simplest model with only two
parameters -- the minimum mass $\mmin$ and the amplitude $\fesca$:

\begin{equation}
\rfesc(M) = \begin{cases}
\fesca, & M>\Mcrit \\
0, & M\leq \Mcrit.
\end{cases}
\label{eq:fesc}
\end{equation}

The mass $\Mcrit$ in Equation (\ref{eq:fesc}) may play multiple roles: it can be the critical mass $\Mcrit \sim 10^{11}\Msun$ from \citep{Gnedin2008a} who found that dwarf galaxies below that mass have very low escape fractions; alternatively, it can be the minimal mass of FZH04 that corresponds to the virial mass of a halo with temperature $10^4\dim{K}$ \citep[the critical temperature for production of ionizing photons,][]{Choudhury2008}, or it can be a minimum mass of a halo capable of retaining photoionized gas \citep{Okamoto2008}.

\subsection{Halo bias}
\label{sub:bias}

The bias factor modifies the halo mass function in the regions with non-zero cosmic over-density $\delta$, and,
therefore, forces the photon production rate to become a function
of local over-density. To calculate the total luminosity 
in a given region, we need to integrate the luminosity-mass relation
$L_{<912\angstrom}(M,\,z)$ from \S\ref{sub:escape}
over the halo mass function to find the average UV luminosity
produced by a collapsed baryon at redshift $z$,
\begin{equation}
 L_{\rm tot}(z,\delta,\ss)=\int L_{<912\angstrom}(M,z)\times\dfrac{dn}{dM}(z,\,\delta,\,\ss)\, dM,
\label{eq:Ltot}
\end{equation}
where we use the bubble mass $m$ instead of its $R$ as a variable, as a more convenient variable (in this formulation there is no need to assume a particular geometric shape for a bubble). We adopt the mass function from \citet{Tinker2008} with the halo bias model from \citet{Tinker2010}, modified in the following way to ensure the exact conservation of ionizing photons:
\begin{equation}
\dfrac{\mathrm{d}n}{\mathrm{d}M}=
\dfrac{f_\mathrm{coll}^\mathrm{PS}}{f_\mathrm{coll}^\mathrm{Tinker}}\times
\dfrac{\mathrm{d}n_\mathrm{Tinker}}{\mathrm{d}M}
\end{equation}

The factor
$f_\mathrm{coll}^\mathrm{PS} /
f_\mathrm{coll}^\mathrm{Tinker}$ is required, since the original Press-Schechter form for the halo mass function allows to maintain the exact photon conservation independently of the smoothing scale by virtue of the property
\[
  \langle f_\mathrm{coll}^\mathrm{PS}(\delta,\ss) \rangle = f_\mathrm{coll}^\mathrm{PS}(\delta=0,\ss=0),
\]
while fits provided in \citet{Tinker2010} do not guarantee such normalization. We further discuss the conservation of photons in the FZH04 model in the Appendix.

The Lagrangian bias that is used in \citet{Tinker2010} works only on large scales (small $\ss$) and small over-densities. In ideal case we should know local non-linear Eulerian bias. Such information might be extracted from large scale simulations, like it was done in \citet{Roth2011}. Unfortunately, at the moment these models can't cover all the redshifts and scales that are considered in our reionization model.

\subsection{Lyman Limit Systems}
\label{sub:lls}

In the original FZH04 model the universe remains 100\% ionized after the end of reionization. The mean free path of an ionized photon in such a universe would be limited by the cosmic horizon only. In reality, however, the mean free path for ionizing radiation is much shorter than the cosmic horizion even at lower redshifts \citep{Songaila2010}; it is limited by absorption in the Lyman Limit systems. Hence, the Lyman Limit systems are not modeled by the original FZH04 model and need to be accounted for separately. 

If the mean free path is much larger than the size of an ionized bubble, the majority of photons produced inside a bubble will be able to reach the bubble edge and ionize fresh neutral hydrogen outside the bubble. In the opposite case, when the mean free path is small compared to the bubble size, the probability of absorption by Lyman Limit systems is much higher and only galaxies near the edge of the bubble contribute to ionizing fresh neutral hydrogen beyond the edge. In Appendix we calculate the fraction $f_{LLS}(z,R)$ of ionizing photons emitted inside the bubble that are available for ionizing fresh neutral hydrogen (and, hence, contributing to $\dotNgH(z,\delta,R)$). For such a calculation, distributions of galaxies and Lyman Limit systems inside every bubble is needed; such distributions are not part of the Press-Schechter formalism and, hence, have to be added as external assumptions in the model. In this paper we consider two limiting cases: a homogeneous distribution (galaxies are distributed uniformly inside a bubble) and a highly clumped distribution (all galaxies are located at the bubble center). Lyman Limit systems in both cases are assumed to be uniformly distributed inside the bubble.

The average abundance of Lyman Limit systems is not a free parameter, however; it has been constrained observationally up to $z\approx6$ \citep{Songaila2010}. For our purpose it is more convenient to use the mean free path due to Lyman Limit system, $l_{\rm LLS}$, directly, hence we adopt Equation (8) from \citet{Songaila2010}:
\[
l_{\rm LLS} (z) = 50\left[\dfrac{1+z}{4.5}\right]^{-4.44} \rm Mpc.
\]
This fit is made in the redshift interval $0-6$. In our model we extrapolate this fit to higher redshifts, where its accuracy cannot be verified, but, at present, there is no alternative.

To accurately calculate the effective fraction of photons due to Lyman Limit systems
at some redshift we should know not only the bubble sizes but
also the merging statistics of these bubbles. For instance, if we have
a large number of small bubbles that merge into larger ones only during the
late stages of reionization, 
the effective fraction will be always close to unity. In the opposite case, if all
galaxies are highly clustered at the centers of a few large bubbles, the bubble size will be limited by the 
mean free path and a large fraction of ionizing photons will be lost in Lyman Limit systems.
Luckily, the FZH04 model contains the information about bubble merging.

\subsection{Merging of bubbles}
\label{sub:merging}

The procedure of searching the progenitors using excursion 
set formalism was first described in \citet{Lacey1993} for
the Press-Schechter model. Here we follow the same idea.
It was already applied to the FZH04 model in \citet{Furlanetto2005}, where 
the distribution of bubble progenitors was found. 
However, the authors did not use this information for correcting the number of ionizing photons.

\begin{figure}
\includegraphics[width=1\columnwidth]{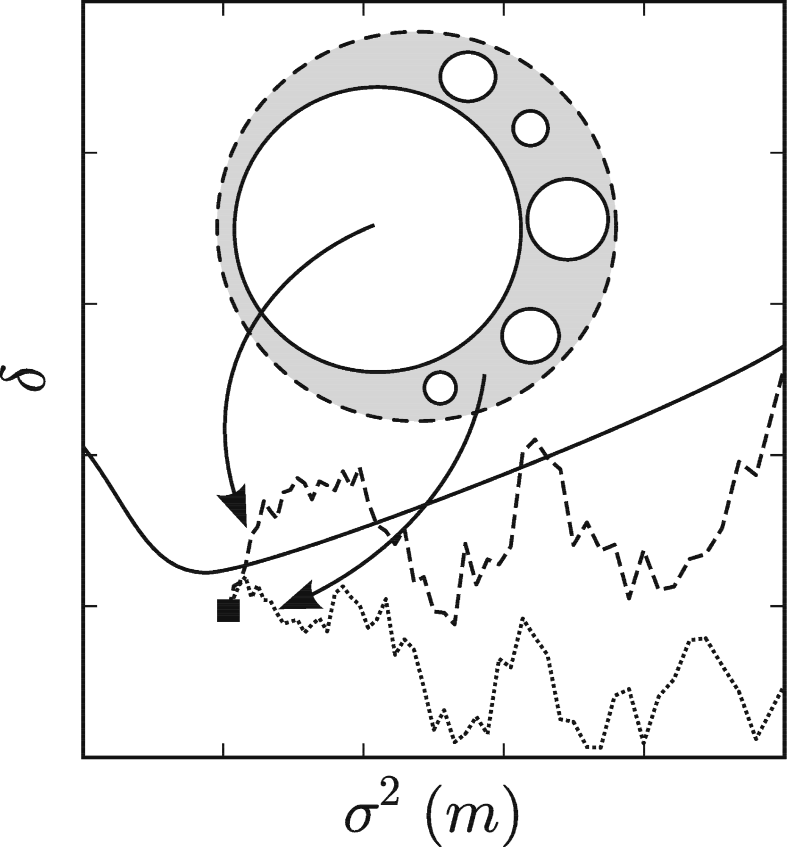}
\caption{Illustration of a barrier at time $t$ (solid line) and two random walks (dashed and dotted lines) started
at some point (square) which belongs to the barrier at time $t+\Delta t$. The dashed random walk crosses
the barrier and, therefore, corresponds to the region of space that was already ionized at time $t$. The dotted
random walk corresponds to the region which was ionized in the time interval between $t$ and $t+\Delta t$.}
\label{fig:random}
\end{figure}

Let's imagine that at some time $t$ we have a distribution of ionized bubbles
in space and assume that there is a barrier that can describe this distribution
\footnote{This assumption is critical in this theory.}. 
During some time $\Delta t$ each of the bubbles was growing
according to the photon production rate from Equation (\ref{NgammaH2}), as well as merging with other bubbles.
In order to calculate $\NgH$ for a bubble
of size $m$ at $t+\Delta t$ we need to know its progenitors, which is illustrated in Figure \ref{fig:random}. A solid line corresponds to the
barrier at time $t$. If we start a random walk at some point below
the barrier (a black square) assuming that this point belongs to the new barrier
at $t+\Delta t$, the trajectories that cross the barrier at time $t$ will correspond to the 
progenitors of the current bubble. Two random walks are shown in the figure: the dashed one crosses the barrier and therefore corresponds to the ionized
bubble at time $t$; the dotted random walk does not cross the barrier, thus it
corresponds to the region that was neutral at time $t$ and got ionized during the time interval between $t$ and $t+\Delta t$.

If we know the distribution $f(\ss(m))$ of the barrier crossings at time $t$, we can write $\NgH$ as: 
\begin{eqnarray}
\label{eq:superN}
 & \NgH(t+\Delta t,\delta',m')\times m'(1+\delta')=\nonumber \\
 & \int_{\ss(m')}^{\ss(\mmin)}f(\ss(m))\times m(1+\delta)\times\nonumber \\
 &  \left(\NgH(t,\delta,m)
 +\dotNgH(t,\delta,m)\Delta t\right)\, d\ss(m),
\end{eqnarray}
where we took into account the initial numbers of photons in each bubble
and their growth during $\Delta t$. The function  $f(\ss(m))$ in Equation (\ref{eq:superN})
depends on the barrier at time $t$.
Solving Equation (\ref{eq:superN}) for every $m'$ gives us the barrier at time $t+\Delta t$.

The described approach makes the model more computationally expensive than the original FZH04 approach, but still much faster than a full numerical simulations.

\begin{figure}[!t]
\includegraphics[width=1\columnwidth]{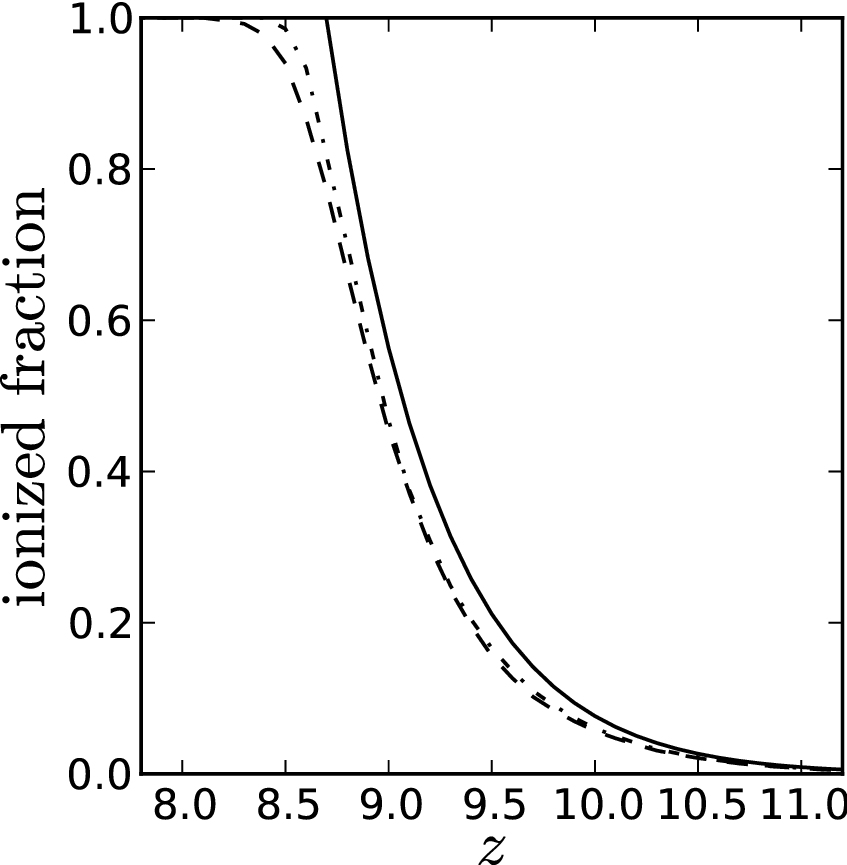}
\caption{\label{fig:barriers}Three reionization histories for a representative model for the escape fraction ($\Mcrit=3\times10^{10} M_{\Sun}$ and $\fesca=0.4$). Solid line -- without LLS; dashed line -- with assumption that galaxies are uniformly distributed inside ionized bubbles; dash-dot line -- with assumption that galaxies are clumped in the center of ionized regions and LLS are uniformly distributed.
}
\end{figure}
\begin{figure}
\includegraphics[width=1\columnwidth]{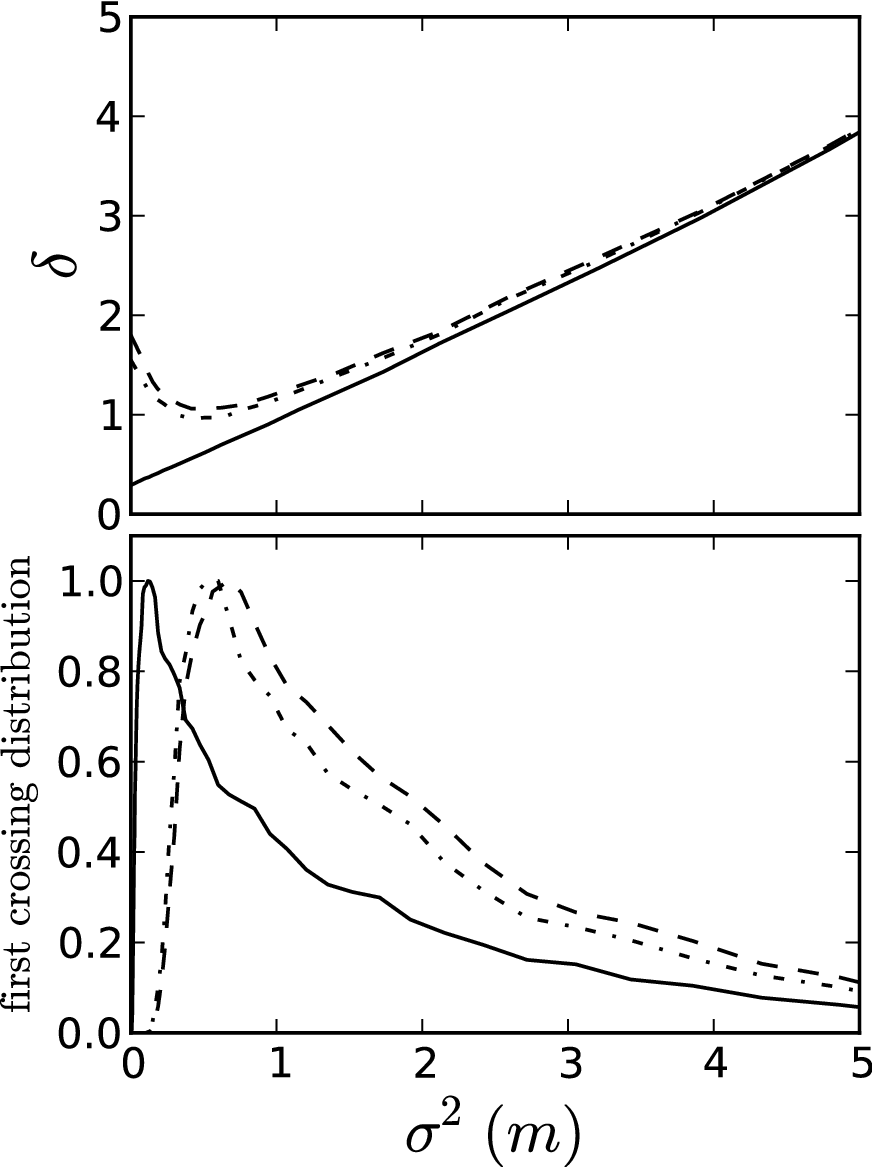}
\caption{\label{fig:biasbarriers}
Barriers (top panel) and first crossing distributions (bottom panel) at $z=9$ for the three models shown in Fig.\ \ref{fig:barriers}.}
\end{figure}

\section{Results}
\label{sec:results}

\begin{figure}[t]
\includegraphics[width=1\columnwidth]{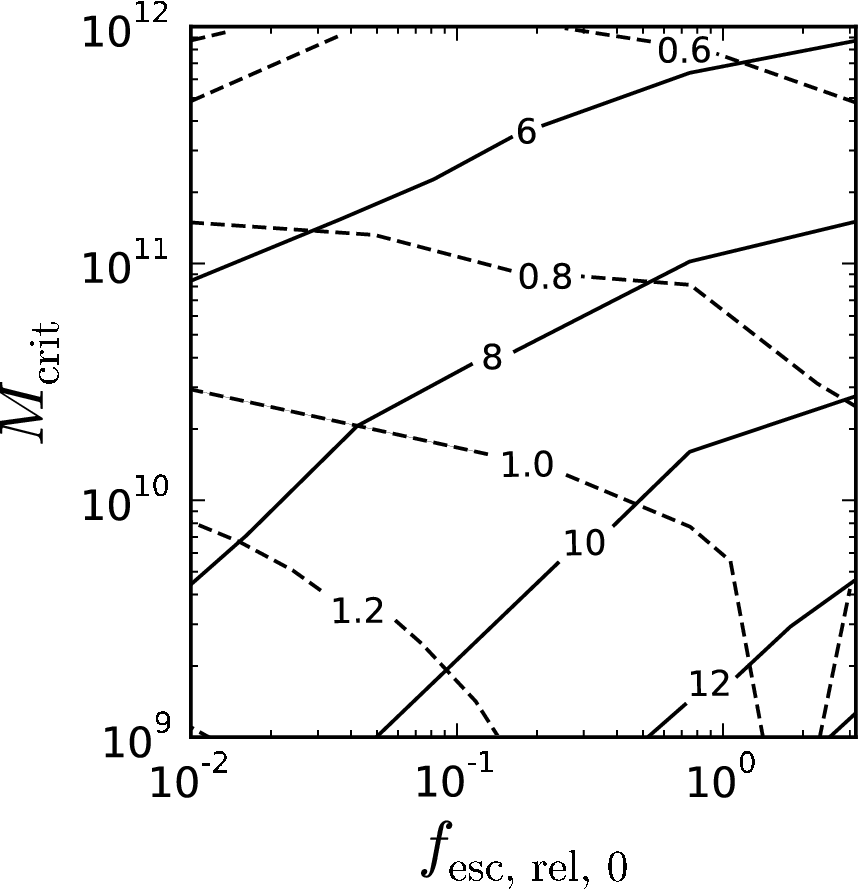}
\caption{\label{fig:zend99}End of reionization (measured as 99\% ionization level, solid lines) the duration of reionization (redshift interval between 20\% and 99\% level of ionization, dashed lines) in the parameter space ($\Mcrit$ and $\fesca$) of our reionization model.}
\end{figure}

\begin{figure}[t]
\includegraphics[width=1\columnwidth]{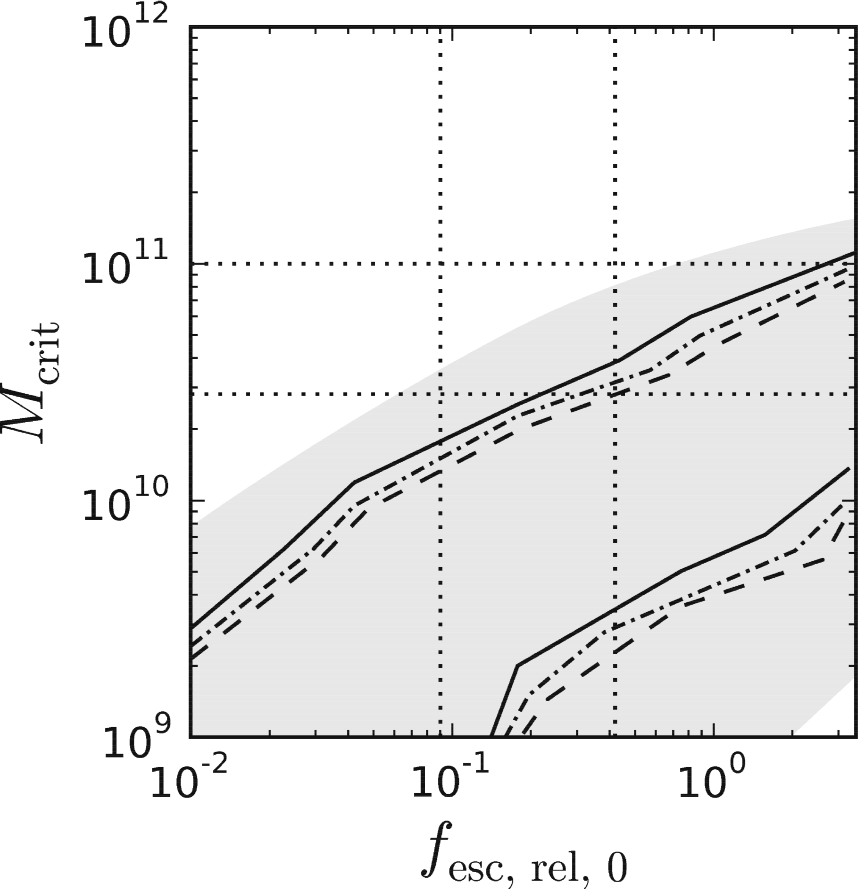}
\caption{\label{fig:Mcritfesc}
Observational constraints in the parameter space ($\Mcrit$ and $\fesca$) of our reionization model (which, \emph{by construction}, matches the observed evolution of the galaxy UV luminosity function for $z\la 8$ and the observed opacity of the Ly-$\alpha$ forest at $z \la 6$). Solid, dot-dashed, and dashed lines show the WMAP+SPT constraint on the Thompson optical depth \citep{Keisler2011} for models without LLS, with LLS and uniformly galaxy distribution, and with LSS and highly clustered galaxy distribution respectively. The shaded region is the SPT constraint on the duration of reionization from \citet{Zahn2012}. For reference, we also show with the dotted horizontal lines the observed constraint on the relative escape fraction at $z\sim3$ from \citet{Nestor2012} and with the vertical dotted lines the constraint on the critical mass from \citet{Fernandez-Soto2003} (see text for detail). Constraints on the duration of reionization from EDGES \citep{Bowman2012EDGES} are, presently, too weak to be able to exclude any portion of the displayed parameter space.}
\end{figure}

The significance of Lyman Limit systems is demonstrated in Figure \ref{fig:barriers}, where we show three reionization histories with Lyman Limit systems effects turned on and off. The corresponding barriers at $z=9$ are shown in the top panel of  Figure \ref{fig:biasbarriers}. In the bottom panel the first crossing time distribution for the barriers are presented. Additional absorption by Lyman Limit systems leads to the upturn of the barrier on large scales (or small $\ss(m)$), because in large ionized bubbles only galaxies near the edge contribute to the further expansion of a bubble. Lyman Limit systems affect the reionization history at later stages, when the average bubble size is large, significantly slowing down the reionization process and making the end of reionization more gradual.

The halo bias modifies the halo mass function, resulting in higher abundance of massive halos in over-dense regions, and, consequently, a higher photon production rate. That lowers the barrier on all scales. However, the effect caused by bias is minor compared to the one due to Lyman Limit systems. Note that the bias effect, in contrast to Lyman Limit systems, does not change the total number of ionizing photons, and, therefore, does not affect the average history of reionization.

The computational efficiency of this method allows us to explore the parameter space of our model for the escape fraction (Equation \ref{eq:fesc}). The results are presented in the Figure \ref{fig:zend99}. Solid contour lines show the end of reionization for the model without Lyman Limit systems, since the end of reionization in models with Lyman Limit systems is less defined (because the end of reionization is gradual). Dashed contour lines show the corresponding duration of reionization, the redshift interval between 20\% and 99\% levels of ionization (following the definition from \citet{Zahn2012}). 

The duration of reionization ($\Delta z$) varies from $0.4$ to $1.6$ across the parameter space presented in the figures. Current constraints on $\Delta z$ from CMB are too broad and can not rule out any of our model parameters. The lower limit on $\Delta z$ made by EDGES \citep{Bowman2012EDGES} and is $\rm \Delta z > 0.06$ with 95\% confidence level, which is much smaller than any of our predictions. 

Other observational constraints in our parameter space ($\Mcrit$ and $\fesca$) are shown in Figure \ref{fig:Mcritfesc}. For all our reionization models consistent with the WMAP value for the Thompson optical depth \citep[$\tau = 0.085\pm 0.014$][]{Keisler2011}, the measurement of the secondary anisotropies by South Pole Telescope \citep[SPT][]{Zahn2012} does not yet provide any significant constraint, because reionization proceeds faster than the SPT upper limit on the duration of reionization. We also show, mostly for the sake of illustration, in Figure \ref{fig:Mcritfesc} observational constraints on out two main parameters ($\Mcrit$ and $\fesca$) obtained at lower redshifts $z\sim3$: the actual measurement of $\fesca$ by \citet{Nestor2012} and the constraint on $\Mcrit$ by \citet[][if we interpret their non-detection of the escape of ionizing radiation for the low-luminosity sub-sample as a cutoff in the escape fraction for low mass galaxies]{Fernandez-Soto2003}. We reemphasize that the latter two constraints are obtained at $z\sim3$ and, therefore, may not apply to galaxies at $z>6$.


\section{Conclusions}
\label{sec:conclusions}

In this paper we extend the analytical model for cosmic reionization from FZH04 to account for the absorption of ionizing radiation by Lyman Limit systems, for the halo bias effect, and for merging of ionized bubbles.

The halo bias doesn't affect the average reionization history since it doesn't change the total number of galaxies (and, hence, the total number of ionizing photons). However, the halo bias causes a redistribution of ionizing photons between over-dense and under-dense regions, and, therefore, modifies the size distribution of ionized bubbles by increasing the abundance of large bubbles and suppressing small ones.

Additional absorption of ionizing radiation by Lyman Limit systems is significant only during the late stages of reionization, when sufficiently large bubbles (with sizes exceeding the photon mean free path due to Lyman Limit systems) are abundant.

Our model is, by construction, consistent with the observed evolution of the galaxy luminosity function at $z\la8$ and with the observed evolution of Ly-$\alpha$ forest at $z\la6$. We also find that, for a wide range of values for the relative escape fraction that we consider reasonable, and which are consistent with the observational constraints on the relative escape fraction from lower redshifts, our reionization models are consistent with the WMAP constraint on the Thompson optical depth and with the SPT and EDGES constraints on the duration of reionization. We, therefore, conclude that it is possible to develop physically realistic models of reionization that are consistent with all existing observational constraints.

\acknowledgements 

This work was done with significant usage of CosmoloPy Python package\footnote{http://roban.github.com/CosmoloPy/}.

This work was
supported in part by the DOE at Fermilab, by the NSF grant
AST-0908063, and by the NASA grant NNX-09AJ54G. The simulations used
in this work have been performed on the Joint Fermilab - KICP
Supercomputing Cluster, supported by grants from Fermilab, Kavli
Institute for Cosmological Physics, and the University of Chicago.
This work made extensive use of the NASA Astrophysics Data System and
{\tt arXiv.org} preprint server.


\bibliographystyle{apj}
\bibliography{refs,igm,self,dsh,rei}


\appendix

\section{Galaxy spectrum}

In this appendix we compute the fraction photons produced by galaxy that is
 energetic enough to ionize hydrogen. We use the simple fit to
the stellar spectrum produced by Starburst99. The Equation 2 in \citet{GnedinHollon2012} \citep[Figure 4 in][]{Ricotti2002}:

\begin{equation}
x\equiv\dfrac{h\nu}{1\,\mathrm{Ry}}
\end{equation}

\begin{equation}
s_{\nu}(\nu)=\dfrac{1}{5.5}\begin{cases}
5.5, & x<1\\
x^{-1.8}, & 1<x<2.5\\
0.4x^{-1.8}, & 2.5<x<4\\
2\times10^{-3}x^{3}/(\exp(x/1.4)-1), & 4<x
\end{cases}
\end{equation}

Since the observations of galaxies \citep{Bouwens2007,Bouwens2010} provide us with information about the luminosity at wavelength $\lambda=1600A$, we want to convert it into ionization luminosity (wavelengths $<912A$):

\begin{eqnarray}
L_{\gamma}^{int}(<912)=\alpha\,\nu L_{\nu}^{obs}(1600)\label{eq:Lintgamma},
\end{eqnarray}
where $\alpha$ is the constant that converts observed value $\nu L_{\nu}^{obs}(1600A)$ into total flux $L_{\gamma}^{int}(<912A)$. Let's find $\alpha$. We know that $L_{\nu}\sim s_{\nu}$. So:

\begin{equation*}
\alpha\times(x_{\nu}s_{\nu}(hc/1600A/1Ry))=\int_{1}^{\infty}s_{\nu}dx,
\end{equation*}
from which we find:
\begin{equation}
\alpha=0.241\label{eq:alphaLgamma}.
\end{equation}

The average energy of ionizing photon can be calculated as

\begin{equation}
\left\langle E\right\rangle =\dfrac{\int_{1}^{\infty}h\nu\times s_{\nu}dx}{\int_{1}^{\infty}s_{\nu}dx}=1.93\, Ry\label{eq:averageE},
\end{equation}

and the emission rate of ionizing photons is 

\begin{equation}
\label{eq:Ndot}
\dot{N_{\gamma}} =\alpha\nu L_{\nu}^{obs}(1600\angstrom)/\left\langle E\right\rangle.
\end{equation}

\section{Fraction of escaped photons}
Here we calculate the fraction of photons that end up in Lyman Limit systems and the fraction that successfully reach the boundary of an ionized bubble
and effectively excite the baryon on the edge. 

Firstly we consider the case when galaxies and LLS are distributed uniformly inside the bubbles.
When the mean free path (MFP) is larger than bubble size the number of effective photons
is proportional to the volume. Another regime is when MFP is much
smaller, in this case the number of photons is proportional to its
surface area.

If the galaxies and Lyman Limit systems are distributed
uniformly in the ionized sphere, what fraction photons can 
reach the surface of the bubble? Assume that MFP at that epoch is
$l_\mathrm{LLS}$.

Consider a bubble of size R and a galaxy located at distance $h$ from
the center. The fraction of photons that will reach the boundary
of the bubble will be:
\[
\sqrt{h^{2}+R^{2}-2Rh\cos(\theta)}=x,
\]
\[
f_\mathrm{eff}(h)=\dfrac{1}{4\pi R^{2}}\int_{0}^{\pi}d\theta2\pi\sin(\theta)\times x^{2}\times\exp(-x/l_\mathrm{LLS}).
\]

Now we have to find the average over volume:
\[
f_\mathrm{LLS}(R/\lambda)=\dfrac{1}{\frac{4}{3}\pi R^{3}}\int_{0}^{R}4\pi h^{2}f_\mathrm{eff}(h)dh.
\]

In result we have the $f_\mathrm{LLS}$ as a function of $R/l_\mathrm{LLS}$. Notice, that $l_\mathrm{LLS}$ is the function of redshift.

The second case is when galaxies are clumped in the centers of bubbles, but LLS are still distributed uniformly. In this case the fraction of not-absorbed photons is simple:
\[
f_\mathrm{LLS}(R/l_\mathrm{LLS})=1-\exp(-R/l_\mathrm{LLS})
\]

\section{Conservation of photons}

In the original FZH04 model the number of photons is not conserved; i.e.\ the ionized fraction of the universe calculated with the excursion set approach is systemically below the ionized fraction calculated directly, by counting all emitted photons (i.e.\ $\zeta f_\mathrm{coll}(\sigma^2=0)$). This non-conservation is due to trajectories that never cross the barrier - these trajectories correspond to spatial locations that are never fully ionized internally, but some ionizing photons are still emitted inside them (there are just not enough of them). These photons are not accounted for in the original FZH04 model. To correct for that non-conservation, we extended the original FZH04 barrier from Equation (\ref{eq:barrier}) with an additional vertical barrier at $\sigma^2(M_\mathrm{min}+\epsilon)$, where $\epsilon$ is small (the barrier is undefined if $\epsilon=0$). Each trajectory that crosses the vertical part of the barrier has some collapsed fraction at this smoothing scale and, therefore, the ionized fraction of corresponding region is $\zeta f_\mathrm{coll}(\sigma^2(M_\mathrm{min}+\epsilon))$, which is less then unity but larger than zero. Taking into account all these regions, we restore the exact photon conservation in the original FZH04 model and in our extension to it.
\end{document}